\documentclass[a4paper,10pt]{article}
\pdfoutput=1  

\usepackage{jheppub}  
\usepackage{amsmath}
\usepackage[T1]{fontenc}  
\usepackage{graphicx}
\usepackage{epstopdf}
\usepackage{bm}
\usepackage{epsfig}
\usepackage{graphics}
\usepackage{xspace}
\usepackage{hyperref}
\usepackage{slashed}
\usepackage{xcolor}
\usepackage{longtable}
\usepackage{verbatim}
\usepackage[small]{subfigure}
\usepackage[normalem]{ulem}
\usepackage[utf8x]{inputenc}

\newcommand{\Tr}{\text{Tr}}
\newcommand{\Li}{\text{Li}}
\newcommand{\ot}{\leftarrow}

\newcommand{\nn}{\nonumber}

\newcommand{\be}{\begin{equation}}
\newcommand{\ee}{\end{equation}}
\newcommand{\bea}{\begin{eqnarray}}
\newcommand{\eea}{\end{eqnarray}}
\newcommand{\balign}{\begin{align}}
\newcommand{\ealign}{\end{align}}

\newcommand{\bg}{\begin{gather}}
\newcommand{\foma}{\end{gather}}
\newcommand{\noopsort}[1]{}

\newcommand{\vecb}[1]{\mbox{\boldmath $#1$}}

\renewcommand{\vec}[1]{\bm{#1}}

\def\E{\hbox{$\cal E $}}

\def\<{\langle}
\def\>{\rangle}

\def\d{\delta}

\def\({\left(}
\def\[{\left[}
\def\){\right)}
\def\]{\right]}

\def\ln{\hbox{ln}}

\def\Tr{\hbox{Tr}}

\usepackage[normalem]{ulem} 
\renewcommand\sout{\bgroup \color[rgb]{1,0,0} \ULdepth=-.5ex \ULset}

\title{Transverse momentum dependent transversely polarized distributions at next-to-next-to-leading-order}

\author[a]{Daniel Gutierrez-Reyes,}
\author[a]{Ignazio Scimemi,}
\author[b]{and Alexey Vladimirov}

\affiliation[a]{Departamento de F\'isica Te\'orica, Universidad Complutense de Madrid (UCM),\\E-28040 Madrid, Spain}
\affiliation[b]{Institut f\"ur Theoretische Physik, Universit\"at Regensburg, D-93040, Regensburg, Germany}

\emailAdd{dangut01@ucm.es}
\emailAdd{ignazios@fis.ucm.es}
\emailAdd{alexey.vladimirov@ur.de}

\abstract{We calculate the matching of the transversity and pretzelosity transverse momentum dependent  distributions (TMD) on transversity collinear distribution at the next-to-next-to-leading order (NNLO). We find that the matching coefficient for pretzelosity distribution is zero, despite the matrix element for it is nontrivial. This result suggests that the pretzelosity matches a twist-4 distribution. The matching for transversity TMD distributions is provided for both parton distribution functions and fragmentation functions cases.}

\begin{document} 
\maketitle
\flushbottom

\section{Introduction}

The transverse momentum dependent  distributions (TMD) offer the possibility to study the structure of hadrons in great  detail~\cite{Angeles-Martinez:2015sea}. The TMD parton distribution functions (TMDPDF) and fragmentation functions (TMDFF) enter factorization theorems for cross sections of such processes as Drell-Yan or semi-inclusive deep inelastic scattering (SIDIS)~\cite{Collins:2011zzd,Echevarria:2012js,Echevarria:2014rua,Vladimirov:2017ksc}. The usage of the highest available perturbative input is important for the successful description of the experimental data, and significantly increases the predictive power of the framework \cite{Scimemi:2017etj,Scimemi:2018xaf}. Recently, many efforts were made to evaluate elements of TMD factorization theorem at NNLO. As a consequence, the evolution of TMD distributions has been calculated  at two and three loops~\cite{Echevarria:2015byo,Li:2016ctv,Vladimirov:2016dll} and the matching of unpolarized TMDPDFs and TMDFFs have been calculated up to two loops  respectively in \cite{Catani:2011kr,Catani:2012qa,Catani:2013tia,Gehrmann:2012ze,Gehrmann:2014yya,Echevarria:2016scs} and in \cite{Echevarria:2015usa,Echevarria:2016scs}. The status of the polarized distributions is weaker, given also a large number of different distributions. In this work, we consider transversity and pretzelosity (also called quadrupole) TMD distribution, and we evaluate their twist-2 matching at  two loop order.

Both these distributions have been recently subject of experimental, phenomenological and theoretical investigations. The SIDIS data relevant for this extractions come  mainly from HERMES ~\cite{Airapetian:2010ds} and COMPASS~\cite{Alekseev:2008aa,Adolph:2014zba}. Recently also RHIC has provided data in this direction~\cite{Adamczyk:2017ynk} and we expect that transversity will be one of the central measurements in future EIC and LHCSpin. The transverse momentum dependent transversity has been extracted using SIDIS data by  Anselmino et al. in~\cite{Anselmino:2007fs,Anselmino:2008jk,Anselmino:2013vqa} with Gaussian models without taking into account the TMD evolution. In refs. \cite{Kang:2014zza,Kang:2015msa} the lowest order evolution has been considered. An issue of these extractions is the size of the theoretical error. The reduction of it essentially requires the inclusion of the higher order perturbative information. The detailed discussion on theoretical errors, their dependence on perturbative order and related issues has been recently produced in \cite{Scimemi:2018xaf}. Let us also mention here the attempts to constrain the transversity distributions by Monte Carlo and lattice collaborations \cite{Lin:2017stx}. More work in this sense is expected in the future. The transversity TMDFF is also an interesting and practically important object, see \cite{Metz:2016swz} for a recent review. For what concern pretzelosity, we mention here the recent analysis made in \cite{Lefky:2014eia,Parsamyan:2018evv}. According to this analysis, the pretzelosity distribution is very small and practically consistent with a null value.

The one loop results for twist-2 TMDPDFs matching have been obtained recently by our group in \cite{Gutierrez-Reyes:2017glx}. The transversity and pretzelosity belong to the set of TMD distributions which in the regime of large transverse momentum or, equivalently, the small transverse distance match on twist-2 functions. This fact allows us to perform the two-loop calculations in a manner similar to \cite{Echevarria:2016scs}. Namely, we start from the operator definition of transversely polarized TMD distribution and consider its quark matrix element perturbatively with successive matching on the collinear matrix element. The main parts of the calculation practically coincide with the unpolarized case, apart from the algebraic structures and several new master integrals. The calculation is done in the $\delta$-regulator scheme introduced in \cite{Echevarria:2015byo}. The structure of ultraviolet and rapidity divergences is independent of polarization, as it is predicted by TMD factorization theorem \cite{Collins:2011zzd,Vladimirov:2017ksc}, and confirmed by the present calculation. Therefore, the renormalization of these divergences is done using the universal TMD soft factor \cite{Echevarria:2015byo} and renormalization constants calculated in the unpolarized case \cite{Echevarria:2016scs}.

We anticipate here that the matching coefficient obtained for pretzelosity distribution is zero both at one and two loops. This result is particularly surprising since in principle this observable is expected to have a twist-2 contribution \cite{Gutierrez-Reyes:2017glx}. As a counterexample, we recall that the linearly polarized gluon TMD, which has the same quadrupole tensor structure has a  matching coefficient different from zero already at one loop order \cite{Gutierrez-Reyes:2017glx}. So, the theoretical result that we obtain is puzzling. Nonetheless, it confirms the phenomenological and experimental analyses done in \cite{Lefky:2014eia,Parsamyan:2018evv}.
 
For the case of transversity, we provide the matching results at next-to-next-to-leading order (NNLO) both for TMDPDF and TMDFF cases. In the case of TMDFF, we also report the NLO expression, which is missing in the literature to our knowledge. Thus, with the result of this work, the transversity distribution is evaluated at the same level of precision as the unpolarized distributions, and it is the first example of NNLO evaluation of polarized distribution in the TMD factorization formalism. Consequently, the phenomenology for related observables can be developed with a similar level of precision.

We provide the essential notation in sec.~\ref{sec:notation}  and more technical details of TMD distributions in sec.~\ref{sec:technique}. The results for transversity and pretzelosity  TMDPDFs are described in sec.~\ref{sec:trans} and~\ref{sec:pretz} and the case of TMDFF is considered in sec.~\ref{sec:trans-frag}.


\section{Transversely polarized TMD distributions}
\label{sec:notation}

The TMD distribution of the transversely polarized quark is
\begin{align}\label{def:TMD_OP_Q}
\Phi^{[i\sigma^{\alpha+}\gamma^5]}_{q\ot h}(x,\vec b)&=\frac{1}{2}\int \frac{d\lambda}{2\pi}e^{-ixp^+\lambda}
\\ &\nn 
\langle P,S|
\bar T\{\bar q(\lambda n+\vec b)\tilde W_{n}^T(\lambda n+\vec b)\}\,i\sigma^{\alpha+}\gamma^5 \, T\{\tilde W_{n}^{T\dagger}(0)q(0)\}|P,S\rangle,
\end{align}
where index $\alpha$ is transverse and $n$ is a light-like vector. The gauge links $\tilde W_n^T(x)$ are rooted at the position $x$ and continue to the infinity along the direction $n$. We use the standard notation for the light-cone components of vector $v^\mu=n^\mu v^-+\bar n^\mu v^++g_T^{\mu\nu}v_\nu$ (with $n^2=\bar n^2=0$, $n\cdot\bar n=1$).  The superscript $T$ in Wilson lines indicates that they are incorporated by an additional transverse link at light-cone infinity, which ensures the gauge invariance of the operator.

The transversely polarized TMD distribution is parameterized in  terms of four TMD parton distribution functions (TMDPDFs), which were originally introduced in  momentum space \cite{Goeke:2005hb,Bacchetta:2006tn}. For our purposes we need the equivalent parametrization in the position space. It reads
\begin{eqnarray}\label{def:TMD_param}
\Phi^{[i\sigma^{\alpha+}\gamma_5]}_{q\ot h}(x,\vec b)&=&s_{T}^\alpha h_1(x,\vec b)
 -i\lambda b^{\alpha}M h_{1L}^\perp(x,\vec b)
 \\\nn &&+i\epsilon_{T}^{\alpha\mu}b_\mu M h_1^\perp(x,\vec b)+\frac{M^2\vec b^2}{2}\(\frac{ g_{T}^{\alpha\mu}}{2}+\frac{b^\alpha b^\mu}{\vec b^2}\)s_{T\mu} h_{1T}^\perp(x,\vec b),
\end{eqnarray}
where $s_T$ is the transverse part of the hadron spin, $\lambda$ is helicity, $M$ is the mass of hadron and $\vec b^2=-b^2>0$. The detailed relation between momentum and position space definitions can be found in \cite{Boer:2011xd,Scimemi:2018mmi}. In this work we consider the functions $h_1$, which is known as TMD transversity distribution, and in the function $h_{1T}^\perp$, which is known as the pretzelosity distribution.

The small-$b$ operator product expansion (OPE) allows the systematic expansion of the TMD operator in  powers of $\vec b$. The operators that are associated  to  the powers of $\vec b$ are then classified by twists. The evaluation of the matrix element of the small-$b$ OPE results into an expression of the form
\begin{eqnarray}\label{OPE_scheme}
\Phi^{[i\sigma^{\alpha+}\gamma^5]}_{q\ot h}(x,\vec b)&=&\sum_f \[C^{\alpha\beta}_{q\ot f;\text{tw-2}}(\vec b)\otimes h^{\beta;\text{tw-2}}_{f\ot h}\](x)
\\\nn&&\qquad\qquad\qquad
+\vec b_\beta\sum_f\[C^{\alpha\beta\gamma}_{q\ot f;\text{tw-3}}(\vec b)\otimes h^{\gamma;\text{tw-3}}_{f\ot h}\](x)+...~,
\end{eqnarray}
where $h$ are collinear distributions, $C$ are coefficient functions and $\otimes$ is the integral convolution in the momentum fractions. The terms in eq.~(\ref{OPE_scheme}) incorporate all tensor structures of the TMD distribution parametrization in eq.~(\ref{def:TMD_param}). Extracting particular tensors, one can find the matching of individual TMDPDFs onto collinear functions. In particular, the tensor structure of $h_1$ and $h_{1T}^\perp$ appears in the twist-2 term, while the tensor structure for $h_{1L}^\perp$ and $h_1^\perp$ can be produced  only at the twist-3 \cite{Scimemi:2018mmi}. In this work we concentrate on the twist-2 contributions. 

The OPE in transversely  polarized case has exceptionally simple structure, since there are no gluon operators. The only PDF that contribute to the matching of these twist-2 distributions is the collinear  transversity PDF. Its expression reads \cite{Jaffe:1991ra}
\begin{eqnarray}
s_T^\alpha h_1(x)&=&\frac{1}{2}\int \frac{d\lambda}{2\pi}e^{-ixp^+\lambda}
\langle P,S|
\bar T\{\bar q(\lambda n)[\lambda n,0]\,i\sigma^{\alpha+}\gamma^5 \,q(0)\}|P,S\rangle.
\end{eqnarray}
The PDF $h_1(x)$ can also be interpreted as the probability distribution to find  a transversely polarized quark in a hadron. 

The coefficient functions of the OPE are dimensionless and  the dependence on $\vec b$ enters only via logarithms, or via dimensionless tensors. Generally, the twist-2 coefficient functions can have structures $\sim g_T^{\alpha,\beta}$ and $\sim b^\alpha b^\beta/\vec b^2$. It is natural to decompose it as 
\begin{eqnarray}\label{def:coef_tensor}
C^{\alpha\beta}_{q\ot f;\text{tw-2}}(x,\vec b)=g_T^{\alpha\beta}\delta C_{q\ot f}(x,\mathbf{L}_\mu)+\(\frac{g_T^{\alpha\beta}}{2(1-\epsilon)}+\frac{b^\alpha b^\beta}{\vec b^2}\)\delta^\perp C_{q\ot f}(x,\mathbf{L}_\mu),
\end{eqnarray}
here $\epsilon$ the parameter of dimension regularization ($d=4-2\epsilon$), and
\begin{eqnarray}
\mathbf{L}_\mu=\ln\(\frac{\mu^2 \vec b^2}{4e^{-2\gamma_E}}\).
\end{eqnarray}
The  pieces of this decomposition do not mix. In particular, the tensor in the second term 
of eq.~(\ref{def:coef_tensor})
 has zero trace (in $d=4-2\epsilon$ dimensions). Comparing the parametrization in eq.~(\ref{def:coef_tensor}) with the parametrization for TMD distributions we find that the matching for individual TMDPDFs are
\begin{eqnarray}\label{def:mathcTrans}
h_1^q(x,\vec b)&=&\int_x^1\frac{dy}{y}\sum_{f=q,\bar q}\delta C_{q\ot f}\(\frac{x}{y},\mathbf{L}_\mu\)h^f_1(y)+\mathcal{O}(\vec b^2),
\\\label{def:mathcPretz}
h_{1T}^{\perp,q}(x,\vec b)&=&\frac{2}{\vec b^2 M^2}\int_x^1\frac{dy}{y}\sum_{f=q,\bar q}\delta^\perp C_{q\ot f}\(\frac{x}{y},\mathbf{L}_\mu\)h^f_1(y)+\mathcal{O}(\vec b^2),
\end{eqnarray}
where we explicitly express the Mellin convolution integral. In these formulas we also suppress the scale dependence of functions which is discussed in details in the next section. The sum over flavors runs only over non-singlet combinations, since there are no gluon operator with transverse polarized configuration that could mix with the quarks.

The coefficient functions $\delta C$ and $\delta^\perp C$ can be evaluated perturbatively. Naturally, at  tree order $\sim a_s^0$, the coefficient function in eq.~(\ref{def:coef_tensor}) is proportional to $g_T^{\alpha\beta}$, and thus only $\delta C$ is non-zero. 
The terms proportional to $\sim b^\alpha b^\beta/b^2$ are generated by loop-diagrams and appear already at 
one-loop level~\cite{Gutierrez-Reyes:2017glx}. In ref.~\cite{Gutierrez-Reyes:2017glx} it has been found that the 1-loop contribution to $\delta^\perp C$ is zero in  four  dimensions, but  it can be different from zero in dimensional regularization at order $\mathcal{O}(\epsilon)$. This observation suggests that  potentially  this  contribution  does not vanish at two-loop level.

\section{Evaluation of small-$b$ OPE}
\label{sec:technique}

In this section we discuss technical details of the matching procedure at the twist-2 level. In particular, we present the properties of the operator under renormalization and discuss the  divergences that appear along the calculation. Since these details are universal for all TMDPDFs, we do not specify them, but  instead we use the generic notation $\Phi$, that in the context of this work mean $h_1$ or $h_{1T}^\perp$.

\subsection{Renormalization of TMDPDF}
\label{sec:soft}

The TMD operator has two types of divergences. The ultraviolet (UV) and rapidity divergences. Both these divergences are renormalized by appropriate renormalization constants \cite{Vladimirov:2017ksc}. Consequently, the renormalized, and hence physical, TMDPDF depends on two scales. Traditionally, UV renormalization scale is denoted by $\mu$ and the rapidity renormalization scale is denoted by $\zeta$. The renormalized TMDPDF has the form 
\begin{eqnarray}\label{TMD_ren}
\Phi^{\text{ren}}\!(x,\vec b;\mu,\zeta)
=Z(\mu,\zeta|\epsilon)R(\vec b,\mu,\zeta|\epsilon,\delta)\Phi^{\text{unsub.}}(x,\vec b|\epsilon,\delta),
\end{eqnarray}
where we explicitly show the dependence on regularization parameters. In particular, $\epsilon$ is the parameter of dimensional regularization ($d=4-2\epsilon$) and regularizes UV divergences, which are renormalized by the factor $Z$. The $\delta$ is the parameter of $\delta$-regularization \cite{Echevarria:2015byo,Echevarria:2016scs}, and $R$ is the rapidity renormalization factor. The singularities in $\epsilon$ and $\delta$ cancels in the product of eq.~(\ref{TMD_ren}). The renormalization factors are independent of the Lorentz structure but they change according the parton color representation. The renormalization factors  $Z$ and $R$ are intertwined and it is important to specify in which order the divergences are subtracted. Here we work in the scheme where renormalization of rapidity divergences is made prior to the renormalization of UV divergences. The final result for $\Phi^{\text{ren}}$ is of course independent of the subtraction order.

The renormalization factors are scheme dependent. For the UV renormalization we use the $\overline{\text{MS}}$-scheme. To specify the renormalization scheme for rapidity divergences we recall that within the TMD factorization theorem the rapidity divergences are compensated by the soft factor \cite{GarciaEchevarria:2011rb,Collins:2011zzd,Vladimirov:2017ksc}. The soft factor is defined as following
\begin{align} \label{eq:softf}
S(\vec b) &=
\frac{{\Tr}_\text{color}}{N_c}
\langle 0|\[S_n^{T\dagger} \tilde S_{\bar n}^T \](\vec b)
\[\tilde S^{T\dagger}_{\bar n} S_n^T\](0)|0\rangle,
\end{align}
where $S_n$ and ${\tilde S}_{\bar n}$ stand for soft Wilson lines along $n$ and $\bar n$ (for the precise definition of $W^T$ and $\tilde S^T$ see e.g.\cite{Echevarria:2016scs}). The factors $R$ introduced in eq.~(\ref{TMD_ren}) also renormalize the soft factor, such that the whole factorization expression is finite, see the proof and detailed derivation in~\cite{Vladimirov:2017ksc}. It can be shown that within a properly defined scheme the renormalized soft factor is trivial, i.e.
\begin{eqnarray}
R(b,\mu,\zeta|\epsilon,\delta^+)S(\vec b|\epsilon,\delta^+,\delta^-)R(b,\mu,\bar \zeta|\epsilon,\delta^+)=1,
\end{eqnarray}
where $\delta^\pm$ regularize rapidity divergences in the corresponding direction. In this scheme the rapidity renormalization factor has an exceptionally simple form \cite{Echevarria:2012js,Echevarria:2015byo,Vladimirov:2017ksc}
\begin{eqnarray}\label{R=1/S}
R(b,\mu\zeta|\epsilon,\delta^+)=S^{-1/2}\(\vec b|\epsilon,\frac{\delta^+}{2p^+}\sqrt{\zeta},\frac{\delta^+}{2p^+}\sqrt{\zeta}\).
\end{eqnarray}
Such a scheme is very natural since it does not leave any remnant of soft factor in the factorization theorem, and therefore, coincides with other popular schemes of rapidity renormalization, e.g. with the one suggested in \cite{Collins:2011zzd}. Although here we adopt the $\delta$-regularization notation, these expressions could be translated to other regularization schemes, e.g. the translation dictionary of $\delta$-regularization to the regularization by tilted Wilson lines is given in \cite{Buffing:2017mqm}.

\subsection{Scaling properties}

The dependence on renormalization scales is given by the pair of evolution equations
\begin{eqnarray}\label{RGE1}
\mu^2 \frac{d}{d\mu^2}\Phi^{\text{ren}}(x,\vec b;\mu,\zeta)&=&\frac{\gamma_F(\mu,\zeta)}{2}\Phi^{\text{ren}}(x,\vec b;\mu,\zeta),
\\\label{RGE2}
\zeta \frac{d}{d\zeta}\Phi^{\text{ren}}(x,\vec b;\mu,\zeta)&=&-\mathcal{D}(\mu,\vec b)\Phi^{\text{ren}}(x,\vec b;\mu,\zeta).
\end{eqnarray}
The anomalous dimensions are defined via the corresponding renormalization constants. A detailed study of this system has been recently presented in ref.~\cite{Scimemi:2018xaf}. The values of anomalous dimensions are known up to three-loop order inclusively \cite{Moch:2005id,Gehrmann:2010ue,Vladimirov:2016dll,Li:2016ctv}.

The transversity PDF evolves with the DGLAP kernel evolution equation
\begin{eqnarray}\label{RGE3}
\mu^2 \frac{d}{d\mu^2} h^q_1(x,\mu)&=&\sum_{f=q,\bar q}\int_x^1 \frac{dy}{y}\delta P_{q\ot f}\(\frac{x}{y}\)h^f_1(y,\mu),
\end{eqnarray}
where the kernel $\delta P$ is known up to two loop order \cite{Vogelsang:1997ak,Mikhailov:2008my}. Combining together eq.~(\ref{RGE1},~\ref{RGE2}) with eq.~(\ref{RGE3}) we obtain the evolution properties of the matching kernels. They are
\begin{eqnarray}\label{RGE4}
&&\mu^2\frac{d}{d\mu^2}\delta C_{q\ot f}(x,\mathbf{L}_\mu,\mathbf{l}_\zeta)
\\&&\nn\qquad= \sum_{f'=q,\bar q}\int_x^1 \frac{dy}{y}\delta C_{q\ot f'}\(\frac{x}{y},\mathbf{L}_\mu,\mathbf{l}_\zeta\)
\(\frac{\gamma_V(\mu,\zeta)}{2}\delta_{ff'}\delta(\bar y)-\delta P_{f'\ot f}(y)\),
\\\label{RGE5}
&&\zeta\frac{d}{d\zeta}\delta C_{q\ot f}(x,\mathbf{L}_\mu,\mathbf{l}_\zeta)=-\mathcal{D}(\mu,\vec b)\delta C_{q\ot f}(x,\mathbf{L}_\mu,\mathbf{l}_\zeta),
\end{eqnarray}
where $\mathbf{l}_\zeta=\ln(\mu^2/\zeta)$. The analogous expression holds for $\delta^\perp C$ coefficient function. 
In  perturbation theory, the expression for the coefficient function can be presented as
\begin{eqnarray}
\label{eq:pertTR}
\delta C_{f\ot f'}(x,\mathbf{L}_\mu,\mathbf{l}_\zeta)&=&\sum_{n=0}^\infty a_s^n \sum_{k=0}^{n+1}\sum_{l=0}^n \mathbf{L}_\mu^k\, \mathbf{l}_\zeta^l \, \delta C^{(n;k,l)}_{f\ot f'}(x),
\end{eqnarray}
where $a_s=g^2/(4\pi)^2$. The coefficients $\delta C^{(n;k,l)}$ with $k+l>0$ are  fixed order-by-order   with the help of the renormalization group equations in eq.~(\ref{RGE4},~\ref{RGE5}). The expressions for these coefficients in their generic form up to two-loop are given, e.g. in ref.~\cite{Echevarria:2015usa} (see appendix D.1) or in ref.\cite{Scimemi:2017etj} (see appendix B.2). The explicit expressions for the  transversity function are presented in the supplementary file. Thus, the only non-trivial part to evaluate is $\delta C^{(n;0,0)}$.

\subsection{Evaluation of matching coefficient at NNLO}

In order to find the expression for the matching coefficients, we evaluate the matrix elements with free quark states. The spinor indices of the in/out-going quarks can be contracted with $i\sigma^{-\beta}\gamma^5$, since it is the only non-vanishing spinor structure at twist-2. The resulting tensor diagram can be projected on tensor structures of transversity or pretzelosity. The obtained result is then compared with the matched expressions in eq.~(\ref{def:mathcTrans},~\ref{def:mathcPretz}). Schematically, we deal with the equation 
\begin{eqnarray}\label{eq:tomatch}
\Phi_{f\ot q}(x,\vec b)= \sum_{f'=q,\bar q}\delta C_{f\ot f'}\otimes h_{1;f'\ot q}(x),
\end{eqnarray}
where $h_{1;f'\ot q}(x)$  are the PDF evaluated on  free-quark states. This equation can be solved recursively starting from the first non-zero contribution. 

Evaluating Feynman diagrams we keep the momentum of quark collinear, $p^\mu=p^+\bar n^\mu$. This choice of kinematics significantly simplifies the calculation. In particular, it implies that $\vec b^2$ is the only (Lorentz invariant) scale that is presented in the diagrams. Since the scaleless loop-integrals are zero in the dimensional regularization, many  diagrams vanish. For example, this is the case  of  all pure virtual correction diagrams. Also all loop-integrals contributing to PDF are zero. Thus the only non-zero part of $h_1(x)$ is the renormalization constant. 
That is, the needed expression for PDF evaluated on quarks is a pure singularity, which can be found using  renormalization group equations. It reads
\begin{eqnarray}
h^{[0]}_{f\ot f'}(x)&=&\delta_{ff'}\delta(1-x),\qquad h^{[1]}_{f\to f'}(x)=-\frac{\delta_{ff'}}{\epsilon}\delta P^{[1]}_{f\ot f}(x),
\\\nn
h^{[2]}_{f\ot f'}(x)&=&\frac{1}{2\epsilon^2}\left (\sum_{r={q,\bar q}}\d P^{[1]}_{f\ot r}\otimes \d P^{[1]}_{r\ot f'}+\frac{\beta_0}{2}\d P^{[1]}_{f\ot f'}\right)(x)-\frac{1}{2\epsilon}\d P^{[2]}_{f\ot f'}(x),
\end{eqnarray}
where $\delta P^{[n]}$ are perturbative coefficients of transversity DGLAP kernel at $a_s^n$-order \cite{Vogelsang:1997ak,Mikhailov:2008my}, and $\beta_0$ is the QCD $\beta$-function. Note that $\delta P^{[1]}_{q\ot \bar q}(x)=0$. 

The evaluation of TMD  is made using the $\delta$-regularization, which is described in details in \cite{Echevarria:2015byo,Echevarria:2016scs}. It is a very convenient form of regularization, in particular, because it allows a clear separation of divergences. The outcome of each diagram at NNLO has a generic form
\begin{align}
\text{diag.}=(\vec b^2)^{2\epsilon}\Big(f_1(x,\epsilon)&+\left(\frac{\delta^+}{p^+}\right)^{\epsilon}f_2(x,\epsilon)+\left(\frac{\delta^+}{p^+}\right)^{-\epsilon}f_3(x,\epsilon)
\\\nn &+\ln\(\frac{\delta^+}{p^+}\) f_4(x,\epsilon)+\ln^2\(\frac{\delta^+}{p^+}\) f_5(x,\epsilon)\Big).
\end{align}
where the functions $f_i$ are regulated in the limit $x\to 1$ with $+$-distributions. The second and the third terms here represent the IR divergence. Therefore, the functions $f_2$ and $f_3$ exactly cancel in the sum of all the diagrams (and this fact  can be also traced in the sum of sub-classes of diagrams). The last two terms represent the rapidity diverging pieces and thus the functions $f_4$ and $f_5$ are canceled by the rapidity renormalization factor. Altogether these cancellations serve as a good intermediate check of the computation.

Summing together the diagrams we obtain the un-subtracted expression for TMDPDF on free-quark states. Let us denote it as
\begin{eqnarray}
\Phi_{f\ot f'}^{\text{unsub.}}&=\Phi_{f\ot f'}^{[0]\text{unsub.}}\delta_{ff'}+a_s \Phi_{f\ot f'}^{[1]\text{unsub.}}\delta_{ff'}+a_s^2 
\Phi_{f\ot f'}^{[2]\text{unsub.}}+\mathcal{O}(a_s^3),
\end{eqnarray}
and it is UV and rapidity divergent.
Starting from  eq.~(\ref{TMD_ren}) the renormalization procedure reads (here we omit the suffix $^{\text{ren}}$ for simplicity)
\begin{align}
\Phi^{[0]}_{f\ot f'}&=\Phi^{[0]\text{unsub.}}_{f\ot f'}
\\\
\Phi^{[1]}_{f\ot f'}&=\Phi^{[1]\text{unsub.}}_{f \ot f'}-\frac{S^{[1]}\Phi^{[0]\text{unsub.}}_{f \ot f'}}{2}+\left(Z_q^{[1]}-Z_2^{[1]}\right)\Phi^{[0]\text{unsub.}}_{f \ot f'}
 \label{rentransNNLO}
\\
\Phi^{[2]}_{f\ot f'}&=\Phi^{[2]\text{unsub.}}_{ f\ot f'}-\frac{S^{[1]}\Phi^{[1]\text{unsub.}}_{ f\ot f'}}{2}-\frac{S^{[2]}\Phi^{[0]\text{unsub.}}_{ f\ot f'}}{2}+\frac{3S^{[1]}S^{[1]}\Phi^{[0]\text{unsub.}}_{ f\ot f'}}{8}\nn\\
&+\left(Z_q^{[1]}-Z_2^{[1]}\right)\left(\Phi^{[1]\text{unsub.}}_{ f\ot f'}-\frac{S^{[1]}\Phi^{[0]\text{unsub.}}_{ f\ot f'}}{2}\right)
\nn \\ &
+\left(Z_q^{[2]}-Z_2^{[2]}-Z_2^{[1]}Z_q^{[1]}+Z_2^{[1]}Z_2^{[1]}\right)\Phi^{[0]\text{unsub.}}_{f\ot f'},
\end{align}
where superscript in square brackets indicates the perturbative order of a quantity. In this expression we have $Z=Z_2^{-1}Z_q$ with $Z_2$ the quark-field renormalization constant, and $Z_q$ the TMD renormalization constant. 
The expression for them can be found e.g. in \cite{Echevarria:2016scs}. The soft function up to NNLO in $\d$-regularization is calculated in~\cite{Echevarria:2015byo}.

\section{Matching of transversity TMD distribution at NNLO}
\label{sec:trans}

The evaluation of the transversity matching coefficient is very similar to the evaluation of the unpolarized matching coefficient made in \cite{Echevarria:2016scs}. The main difference, which only simplifies the evaluation, is the absence of the mixing with the gluon operator.

The LO contribution expression for transversity TMDPDF is
\begin{align}
&h^{[0]}_{1f\ot f'}(x)=\delta(1-x)\delta_{ff'}.
\end{align}
Substituting it in eq.~(\ref{eq:tomatch}), we find the LO matching coefficient 
\begin{eqnarray}
\delta C^{[0]}_{f\to f'}=\delta C^{(0;0,0)}_{f\to f'}(x)&=&\delta_{ff'}\delta(1-x).
\end{eqnarray}
Using it as a  starting expression for  iteration we obtain
\begin{eqnarray}
\delta C^{[1]}_{f\ot f'}&=&h_{1;f\ot f'}^{[1]}(x,\vec b)-h_{1;f\ot f'}^{[1]}(x),
\\\label{dC2:mathing}
\delta C^{[2]}_{f\ot f'}&=&h_{1;f\ot f'}^{[2]}(x,\vec b)-\sum_{r}\delta C^{[1]}_{f\to r}\otimes h_{1;r\ot f'}^{[1]}(x)  -h_{1;f\ot f'}^{[2]}(x).
\end{eqnarray}
Some intermediate expressions, such as renormalization constants, un-subtracted and subtracted expressions, can be found in the supplementary \textit{Mathematica} file. Here we present only the final result of the evaluation.

The NLO expression for matching coefficient reads
\begin{eqnarray}\label{res:trans:NLO}
\delta C^{[1]}_{f\ot f'}(x,\vec b)&=&C_F \delta _{ff'}\left(
-\frac{4x\mathbf{L}_{\mu}}{(1-x)_+}+\delta(\bar x)\(-\mathbf{L}_{\mu}^2+2\mathbf{L}_{\mu}\mathbf{l}_\zeta-\zeta_2\)\right),
\end{eqnarray}
where the $(..)_+$-distribution is defined as usual
\begin{eqnarray}\label{plus}
\(f(x)\)_+=f(x)-\delta(1-x)\int_0^1 dy f(y).
\end{eqnarray}
This result agrees\footnote{The calculation made in \cite{Bacchetta:2013pqa} is made in a non-standard $\overline{\text{MS}}$-scheme. And for this reason, the coefficient presented in \cite{Bacchetta:2013pqa} is different from eq.~(\ref{res:trans:NLO}) by $\zeta_2\delta(\bar x)$ term.} with the ones obtained in refs.~\cite{Bacchetta:2013pqa,Gutierrez-Reyes:2017glx,Buffing:2017mqm}. It is easy to see that logarithmic part of eq.~(\ref{res:trans:NLO}) satisfies renormalization group equation eq.~(\ref{RGE4},~\ref{RGE5}). The finite part is
\begin{eqnarray}
\delta C^{(1;0,0)}_{f\ot f'}(x)=-C_F\zeta_2\delta_{ff'}\delta(\bar x).
\end{eqnarray}
Note, that in order to evaluate NNLO matching coefficient one needs the terms suppressed by $\epsilon$, since they interfere with the singularities of the PDF, and produce a non-zero finite and $1/\epsilon$ contribution to eq.~(\ref{dC2:mathing}). The complete expression at all orders of $\epsilon$ can be found in \cite{Gutierrez-Reyes:2017glx}.

The expression for $\delta C^{[2]}$ is lengthy. Since the only non-trivial part is the finite part $\delta C^{(2;0,0)}$ we restrict  to it here. The logarithmic part can be restored using the renormalization group, and in the explicit form it is given in the supplementary \textit{Mathematica} file. At NNLO we have the mixing with anti-quark operator, therefore, the matching is split into two channels
\begin{eqnarray}
\label{C0}
\delta C^{(2;0,0)}_{f\ot f'}(x)&=& \delta_{ff'}\delta C^{(2;0,0)}_{q\ot q}(x)+\delta_{f\bar f'}\delta C^{(2;0,0)}_{q\ot \bar q}(x),
\end{eqnarray}
where
\begin{eqnarray}\label{C1}
\delta C_{q\ot q}^{(2;0,0)}(x)&=& C_F^2\Big\{\delta p(x)\Big[4\Li_3(\bar x)-20 \Li_3(x)-4\ln \bar x\, \Li_2(\bar x)+12\ln x\,\Li_2(x)
\\\nn && \qquad\qquad+2\ln^2\bar x\,\ln x+2\ln\bar x\,\ln^2 x+\frac{3}{2}\ln^2 x+8 \ln x+20\zeta_3\Big]-2\ln\bar x+4\bar x\Big\}
\\\nn &&+C_FC_A\Big\{\delta p(x)\Big[8\Li_3(x)-4\Li_3(\bar x)+4\ln\bar x\,\Li_2(\bar x)-4 \ln x\,\Li_2(x)
\\\nn &&\qquad \qquad-\frac{\ln^3x}{3}-\frac{11}{6}\ln^2 x-\frac{76}{9}\ln x+6\zeta_3-\frac{404}{27}\Big]+2\ln\bar x+\frac{14}{3}\bar x\Big\}
\\\nn &&+C_FN_f\Big\{\delta p(x)\Big[\frac{\ln^2x}{3}+\frac{10}{9}\ln x+\frac{56}{27}\Big]-\frac{2\bar x}{3}\Big\}
\\\nn &&+\delta(\bar x)\Big[C_F^2\frac{5\zeta_4}{4} +C_FC_A\(5\zeta_4-\frac{77}{9}\zeta_3-\frac{67}{6}\zeta_2+\frac{1214}{81}\)
\\\nn && \qquad \qquad
+C_FN_f\(\frac{14}{9}\zeta_3+\frac{5}{3}\zeta_2-\frac{164}{81}\)\Big],
\end{eqnarray}
\begin{eqnarray}\label{C2}
\delta C^{(2;0,0)}_{q\ot \bar q}(x)&=&\(C_F^2-\frac{C_FC_A}{2}\)\Big\{\delta p(-x)\Big[8\Li_3\(\frac{1}{1+x}\)-8\Li_3\(\frac{x}{1+x}\)
\\\nn &&\qquad\qquad+4\Li_3(x^2)-4\ln x \Li_2(x^2)+4 \ln^2 x\ln(1+x)-4 \ln x\ln^2 (1+x)
\\&&\nn \qquad\qquad -\frac{2}{3}\ln^3 x-4\zeta_3\Big]+2\bar x\Big\}.
\end{eqnarray}
Here,
\begin{eqnarray}
\delta p(x)=\frac{2x}{1-x},
\end{eqnarray}
is
the regular part of LO DGLAP kernel. In this expression the singularity at $x\to 1$ is treated as the $(..)_+$-distribution. These expressions are one of the main result of this work. 

It is intriguing to observe that the parts of eq.~(\ref{C1},~\ref{C2}) enclosed by the square brackets literally coincide with corresponding parts for the unpolarized matching coefficient, see \cite{Echevarria:2016scs} eqns. (7.3) and (7.8). In other words, the matching coefficient has the form
\begin{eqnarray}\label{NNLOpattern}
C^{(2;0,0)}(x)= P^{[1]}(x)F_1(x)+F_2(x)+\delta(\bar x)F_3,
\end{eqnarray}
where $P^{[1]}(x)$ is LO DGLAP kernel, for the corresponding PDF. Then we observe that the function $F_1(x)$ and the constant $F_3$ are the same for unpolarized and transversity kernels (for both flavor channels). Such behavior is expected  since the contributions proportional to $1/(1-x)$, as well as $\delta(\bar x)$ contributions, that primary form the LO DGLAP kernel, comes from the diagrams where the quarks interact with the Wilson lines. Such diagrams are insensitive to the polarization structure of the operator. The rest of diagrams are not singular in the limit $x\to 1$, and thus form a regular contribution. For more detailed discussion on the internal structure of transversity kernel see \cite{Mikhailov:2008my}.

\section{Matching of pretzelosity distribution at NNLO}
\label{sec:pretz}

The calculation for the matching of the pretzelosity TMDPDF over the transversity integrated PDF is in principle similar  to the  one  of  the transversity TMDPDF. 
In this case one has a different  projector, see eq.~(\ref{def:coef_tensor}),
\begin{align}
\frac{b^\mu b^\nu}{\vecb b^2}+\frac{g_T^{\mu \nu}}{2(1-\epsilon)},
\label{pretzproj}
\end{align}
to be compared to  $g_T^{\mu \nu}$ used in the transversity calculation.
Moreover the relation
\begin{align}
\sigma^{\mu +}\(\frac{b^\mu b^\nu}{\vecb b^2}+\frac{g_T^{\mu \nu}}{2(1-\epsilon)}\)\sigma^{-\nu}=0.
\label{simplpretz}
\end{align}
allows a simplification of many diagrams. In particular, the diagrams with a non-interacting quark line are exactly zero, according to the expression \eqref{simplpretz}. This feature reduces the number of diagrams that we have to calculate for the pretzelosity TMD distribution. The pretzelosity projector is built as a sum of two terms. The first one is $g_T^{\mu \nu}$ and it is the same as in the transversity calculation. As the topology of the diagrams is the same in both cases the integrals that appear in the calculation of the diagrams are also the same. The second term of the s $b^\mu b^\nu/\vecb{b}^2$ and  this implies new types of master integrals, that has scalar products $(b\cdot q)^2$ in the numerator (here, $q^\mu$ is a loop-momentum). Such structures appears due to the convolution of a generic diagram with open indices with the projector (\ref{pretzproj}).

The small-$b$ expression fo the matching of the pretzelosity distribution is written in a form  equivalent to the transversity case,
\begin{align}
h_{1T, f\ot f'}^\perp(x,\vecb b)&=\sum_{r=q,\bar q, q'}\Big[\delta^\perp C_{f\ot r}(\vecb b)\otimes \delta f_{r\ot f'}\Big](x)+\mathcal{O}(\vecb b^2).
\label{smallbPretz}
\end{align}
Note, that the collinear function in eq.~\eqref{smallbPretz} is the transversity PDF. As in the transversity case, at NLO we have only the quark-to-quark channel, and at NNLO we have quark-to-quark and quark-to-antiquark channels.

Due to eq.~(\ref{simplpretz}), the un-subtracted pretzelosity distribution  is zero at LO, i.e. $\delta^{\perp} \Phi^{[0]} (x)=0$. Consequently, the LO matching coefficient is also zero, i.e. $\delta^\perp C^{[0]}_{q\ot q}(x)=0$. This fact induces a simplification in  the renormalization of the pretzelosity TMDPDF at NLO, demanding the absence of any divergences at this order. Moreover, due to the absence of the tree order collinear counterpart for the matching procedure the pretzelosity is suppressed by $a_s$. As a result, the expression for the matching coefficient is given solely by the one-loop TMD matrix element
\begin{align}
\delta^\perp C^{[1]}_{q\ot q}(x,\vecb b)=\d^\perp\Phi^{[1]}_{q\ot q}(x,\vecb b)=-4C_F \boldsymbol{B}^\epsilon \Gamma(-\epsilon)\bar x \epsilon^2.
\label{coeffpretzNLO}
\end{align}
We see that the obtained matching coefficient is $\epsilon$-suppressed, so, in the limit $\epsilon \to 0$ it is zero, i.e. $\delta^\perp C^{[1]}_{q\ot q}(x,\vecb b)=0$. This result is given in \cite{Gutierrez-Reyes:2017glx}. According to eq.~\eqref{coeffpretzNLO}, the pretzelosity distribution is suppressed numerically however this result does not ensure that a non-trivial coefficient can be obtained at higher orders. 

The nullity of the LO pretzelosity distribution and the $\epsilon$-suppressed behavior of the NLO contribution yields in a simple expression for the renormalized pretzelosity TMDPDF at NNLO
\begin{align}
h_{1T, f\ot f}^{\perp[2]}=\d^{\perp}\Phi^{[2]}_{f\ot f'}-\frac{S^{[1]}\d^\perp\Phi^{[1]}_{f\ot f'}}{2}+\left(Z_q^{[1]}-Z_2^{[1]}\right)\d^\perp\Phi^{[1]}_{f\ot f'}
\label{renpretzNNLO}
\end{align}
In this expression it is important to keep all $\epsilon$-terms of $\d^\perp\Phi^{[1]}_{f\ot f'}$, since they are multiplied by factors with leading $1/\epsilon^2$-behavior (the TMD renormalization factor, and soft factor). Thus, these terms produce $1/\epsilon$ singularities, despite the suppressed behavior of $\d^\perp\Phi^{[1]}_{f\ot f'}$. Naturally, these terms cancel the corresponding ultraviolet singularities of the un-subtracted TMD matrix element. We have also checked the exact cancellation of infrared divergences, and rapidity divergences. It is interesting to trace the distribution of the contributions between diagrams with different color factor. There are four types of contribution
\begin{eqnarray}
\d^{\perp}\Phi^{[2]}_{f\ot f'}=C_F^2 A_{F}+C_F\(C_F-\frac{C_A}{2}\)A_{FA}+\frac{C_FC_A}{2}A_{A}+C_FN_fA_{N}.
\end{eqnarray} 
The contribution $A_N$ is $\epsilon$-suppressed in a similar manner as the one-loop expression. The contribution $A_F$ is canceled by the renormalization factor entirely up terms suppressed in $\epsilon$. Thus, the only non-zero contribution to the TMD matrix element comes from $A_{FA}$ and $A_A$, which we find to be equal up to higher powers of $\epsilon$. So, concluding we have found
\begin{eqnarray}
A_F&=&\frac{S^{[1]}\d^\perp\Phi^{[1]}_{f\ot f'}}{2}-\left(Z_q^{[1]}-Z_2^{[1]}\right)\d^\perp\Phi^{[1]}_{f\ot f'}+\mathcal{O}(\epsilon),
\\
A_{FA}&=&A_{A}+\mathcal{O}(\epsilon),\qquad A_N=\mathcal{O}(\epsilon).
\end{eqnarray}
Therefore, the contribution proportional to $C_A$ disappears from the final expression, despite only these diagram are non-trivial. The resulting TMD matrix element in the pretzelosity channel is proportional to $C_F^2$ only, and it reads
\begin{align}
\label{qqpretz}
&h_{1T, q\ot q}^{\perp[2]}=-4 C_F^2 \left(\bar x\left(3+4\ln \bar x\right)+4x \ln x\right)+\mathcal{O}(\epsilon),\\
&h_{1T, q\ot \bar q}^{\perp[2]}=0.
\label{pretz}
\end{align}

Expanding  eq.~\eqref{smallbPretz} up to order $a_s^2$ we obtain the following expressions for matching coefficient
\begin{align}
\label{qqCNNLOp}
&\delta^\perp C_{q\ot q}^{[2]}(x,\vecb b)=h_{1T,q\ot q}^{[2]}(x, \vecb b)-\[\delta^\perp C_{q\ot q}^{[1]}(\vecb b)\otimes \delta f^{[1]}_{q\ot q}\](x)\\
&\delta^\perp C_{q\ot \bar q}^{[2]}(x,\vecb b)=0.
\label{coeffsNNLOpretz}
\end{align}
The convolution term that appears in eq.~(\ref{qqCNNLOp}) is different from zero because the NLO matching coefficient is $\epsilon$-suppressed but the NLO transversity integrated PDF is $\epsilon$-divergent. So, the result for the convolution term is finite,
\begin{align}
\[\delta^\perp C_{q\ot q}^{[1]}(\vecb b)\otimes \delta f^{[1]}_{q\ot q}\](x)=-4 C_F^2 \left(\bar x\left(3+4\ln \bar x\right)+4x \ln x\right),
\end{align}
which 
 is the same that we get in eq.~(\ref{qqpretz}). Using eq.~(\ref{qqCNNLOp}) we obtain a null value for the NNLO pretzelosity to transversity matching coefficient. So,
\begin{align}
\delta^\perp C_{q\ot f}^{[2]}(x,\vecb b)=0+\mathcal{O}(\epsilon),
\end{align}
where $f=q,\bar q$.

We stress once more that the cancellation that lead to the zero result has a non-trivial structure. Because the topologies of diagrams that contribute to convolution term in (\ref{qqCNNLOp}) and to the TMD matrix element (\ref{qqpretz}) are completely different. In the first case, these are ladder diagrams, while in the second case we have diagrams with tree-gluon vertex and non-planar diagrams. All this  indicates the presence of a not yet understood concept behind these cancellations, and it suggests that such cancellations take a place at higher orders as well. Therefore  we conjecture that
\begin{eqnarray}\label{conjecture}
\delta^\perp C_{q\ot f}(x,\vecb b)=0,
\end{eqnarray}
at all orders of  perturbation theory. To support this conjecture we also performed the calculation of the matching coefficient in the large-$N_f$ approximation using the same approach as in ref.~\cite{Scimemi:2016ffw}. We obtained that the resummed expression is also $\epsilon$-suppressed. It gives additional confirmation of the conjecture in eq.~(\ref{conjecture}).

\section{Matching of transversity  TMD  fragmentation function at NLO and NNLO}
\label{sec:trans-frag}

The transversely polarized TMDFFs can be treated similarly to the case of transversely polarized TMDPDF.Despite the different origin and interpretation of these distributions, their perturbative treatment is analogous. Therefore, in this section we collect only the necessary definitions and  results, and we do not provide the intermediate details. 

The unsubtracted transversity TMDFFs are  defined with  the following hadronic matrix elements,
\begin{align}\nn
\Delta^{[i\sigma^{\alpha+}\gamma^5]}_{q\rightarrow N}(z,\vecb b)&=\frac{1}{4 z N_c}\sum_X\int \frac{d\lambda}{2\pi}e^{-ip^+\lambda/z}
\\\nn&
\times\langle 0|T\[\tilde W_n^{T\dagger}q_j\]_{a}(n\lambda +\vec b)
|P,S;X\rangle i\sigma^{\alpha+}\gamma^5\langle P,S;X|\bar T\[\bar q_i \,\tilde W_n^T\]_{a}(0)|0\rangle,
 \label{def_FF_opsand}
\end{align}
which can be parameterized by the form factor decomposition
\begin{eqnarray}\label{def:TMDFF_param}
\Delta^{[i\sigma^{\alpha+}\gamma_5]}_{q\ot h}(x,\vec b)&=&s_{T}^\alpha H_1(x,\vec b)
 -i\lambda b^{\alpha}M H_{1L}^\perp(x,\vec b)
 \\\nn &&+i\epsilon_{T}^{\alpha\mu}b_\mu M H_1^\perp(x,\vec b)+\frac{M^2\vec b^2}{2}\(\frac{ g_{T}^{\alpha\mu}}{2}+\frac{b^\alpha b^\mu}{\vec b^2}\)s_{T\mu} H_{1T}^\perp(x,\vec b),
\end{eqnarray}
where again $s_T$ is the transverse part of the hadron spin, $\lambda$ is helicity, $M$ is the mass of hadron and $\vec b^2=-b^2>0$. We recall that the  parametrization presented here is valid for produced hadrons with spin-1/2. For the scalar or pseudo-scalar produced hadrons,  the functions, $H_1$, $H_{1L}^\perp$ and $H_{1T}^\perp$ are absent. The transversity TMDFF is represented by the function $H_1$. The matching onto the fragmentation function is done as
\begin{eqnarray}\label{def:mathcTransFF}
H_1^q(z,\vec b)&=&\int_z^1\frac{dy}{y^{3-2\epsilon}}\sum_{f=q,\bar q}\delta \mathbb{C}_{q\ot f}\(\frac{z}{y},\mathbf{L}_\mu\)H^f_1(y)+\mathcal{O}(\vec b^2).
\end{eqnarray}
The factor $z^{2-2\epsilon}$ is added to meet the common normalization of collinear FF function, that is defined as
\begin{eqnarray}\label{def:FF_param}
s_{T}^\alpha H_1(x)&=&
\frac{z^{1-2\epsilon}}{4 N_c}\sum_X\int \frac{d\lambda}{2\pi}e^{-ip^+\lambda/z}
\\\nn& &
\times\langle 0|T\[\tilde W_n^{T\dagger}q_j\]_{a}(n\lambda)
|P,S;X\rangle i\sigma^{\alpha+}\gamma^5\langle P,S;X|\bar T\[\bar q_i \,\tilde W_n^T\]_{a}(0)|0\rangle.
\end{eqnarray}
The evolution kernels for the collinear FFs are known at two loops \cite{Vogelsang:1997ak,Mikhailov:2008my}. 

The main difference in evaluation of TMDFFs in comparison to TMDPDFs is the origin of parton momentum in diagrams, which is incoming in the PDF case, and outgoing in the FF case. Therefore, the expressions for TMDFFs could be obtained by the application of the crossing symmetry $x\to z^{-1}$ at the diagram level. This, however, should be done with caution since there is a brunch cut for $x>1$, which should be transformed into a branch cut for $z>1$. Additionally, one should take into account the $z^{\epsilon}$ factors that are present in definitions of FFs, and that mix in the $\epsilon$-expansions with various contributions. For the detailed discussion on relation between PDFs and FFs see \cite{Vogelsang:2009pj} and references within. To avoid these complications, we re-evaluate the PDF master integrals with $x\to z^{-1}$ and reassemble the final result. The LO matching of transversity TMDFF is elementary
\begin{eqnarray}
\label{eval:LOFF}
\delta\mathbb{C}^{[0]}_{f\to f'}=\delta_{ff'}\delta(\bar z).
\end{eqnarray} 
Therefore, the renormalization properties of the FF case are similar to the case  of PDFs and the matching procedure follows the same pattern as in the unpolarized case \cite{Echevarria:2016scs}. For this reason we skip the details of evaluation and present the final result. 

For the presentation of the NLO and NNLO coefficient functions we introduce again  logarithmic decomposition (see eq.~(\ref{eq:pertTR}))
\begin{eqnarray}
\label{eq:pertTRFF}
\delta \mathbb{C}_{f\to f'}(z,\mathbf{L}_\mu,\mathbf{l}_\zeta)&=&\sum_{n=0}^\infty a_s^n \sum_{k=0}^{n+1}\sum_{l=0}^n \mathbf{L}_\mu^k\, \mathbf{l}_\zeta^l \, \delta \mathbb{C}^{(n;k,l)}_{f\to f'}(z).
\end{eqnarray}
The terms accompanied by logarithms, i.e. with $k+l>0$ can be restored with renormalization group equation. For completeness, we present these lengthy expressions in the attached {\it Mathematica} file. The finite part of the coefficients at NLO is given by 
\begin{eqnarray}
\label{c1frag}
z^2\delta \mathbb{C}^{(1;0,0)}_{f\to f'} (z)= C_F (4\ln z\hspace{0.1cm}\delta p(z)-\d (\bar z)\zeta_2)\delta_{ff'},
\end{eqnarray}
where 
\begin{eqnarray}
\delta p(z)=\frac{2z}{1-z},
\end{eqnarray}
is the LO DGLAP kernel for the transversity FF. At NNLO  we have the same mixing with anti-quark operator. The matching is split into two channels
\begin{eqnarray}
\label{C0frag}
\delta \mathbb{C}^{(2;0,0)}_{f\to f'}(z)&=& \delta_{ff'}\delta \mathbb{C}^{(2;0,0)}_{q\to q}(z)+\delta_{f\bar f'}\delta \mathbb{C}^{(2;0,0)}_{q\to \bar q}(z),
\end{eqnarray}
where,
\begin{eqnarray}
\label{C0z}
z^2 \delta \mathbb{C}^{(2;0,0)}_{q\to q}(z)&=&C_F^2 \bigg\{\delta p(z)\bigg[
40\text{Li}_3(z)-4 \text{Li}_3(\bar z)+4\ln \bar z\text{Li}_2(\bar z)-16\ln z\text{Li}_2(z)-\frac{40}{3}\ln^3 z
\\&&\nn \qquad+18\ln^2 z \ln \bar z-2\ln^2 \bar z\ln z+\frac{15}{2}\ln^2z
-8\(1+\zeta_2\)\ln z-40 \zeta_3 \bigg]\\
&&\nn
\qquad+4 \bar z (1+\ln z)+2z(\ln \bar z-\ln z)\bigg \}\\
&&\nn+C_F C_A \bigg \{\delta p(z)\bigg [4  \text{Li}_3(\bar z)+12  \text{Li}_3(z)-4 \ln \bar z  \text{Li}_2(\bar z)-8 \ln z \text{Li}_2(z)+3\ln^3 z \\
&&\nn \qquad
 -4\ln \bar z \ln^2 z-\frac{11}{6}\ln^2 z-12\zeta_2\ln z+\frac{70}{3}\ln z+2\zeta_3-\frac{404}{27}\bigg]\\
&&\nn\qquad
 +\frac{14}{3}\bar z-2 z\ln \bar z-2(1-2z)\ln z\bigg \}\\
&&\nn+C_F N_f \bigg \{\delta p(z) \bigg [\frac{\ln^2 z}{3}-\frac{10}{3}\ln z+\frac{56}{27}\bigg ]-\frac{2}{3}\bar z \bigg\}+\delta(\bar z) \bigg\{\frac{5}{4}\zeta_4 C_F^2 \\
&&\nn+C_F C_A \bigg[\frac{1214}{81}-\frac{67}{6}\zeta_2+65\zeta_4-\frac{77}{9}\zeta_3\bigg]
+C_F N_f \bigg [-\frac{164}{81}+\frac{5}{3}\zeta_2+\frac{14}{9}\zeta_3\bigg]\bigg\},
\end{eqnarray}
\begin{eqnarray}
\label{C0zz}
z^2\delta \mathbb{C}^{(2;0,0)}_{q\to \bar q}(z)&=\(C_F^2-\frac{C_F C_A}{2}\)\bigg \{\delta p(-z)\bigg[8\text{Li}_3\(\frac{1}{1+z}\)-8\text{Li}_3\(\frac{z}{1+z}\)-4\text{Li}_3\(z^2\)\\
&\nn +16 \ln z \text{Li}_2\(z\)-4\ln z  \text{Li}_2\(z^2\)-4\ln z \ln^2 (1+z)\\
&\nn-12\ln^2 z \ln (1+z)+6\ln^3 z+4\zeta_3\bigg]+2\bar z\bigg \}.
\end{eqnarray}
The singularity at $z\to1$ is understood as a $(..)_+$-distribution (\ref{plus}). Similarly to the PDF case, the expressions enclosed by  square brackets in eq.~(\ref{C0z},~\ref{C0zz}) literally coincide with the ones  of  the unpolarized fragmenting function  case, (see  eq. (7.11) and (7.17) in \cite{Echevarria:2016scs}). In other words, it can be written the form (\ref{NNLOpattern}), and the functions $F_1(z)$ and $F_3$ coincide for polarized and unpolarized cases.

We have not evaluated the pretzelosity TMDFF, because its calculation is rather involved. However, we have no doubts that it matching coefficient is zero at NNLO alike the matching of pretzelosity TMDPDF.

\section{Conclusions}
\label{sec:conclusions}
In this work, we have studied the twist-2 matching of transversity and pretzelosity (or quadrupole) TMD distributions. We have derived the matching coefficients for these distributions at next-to-next-to-leading order (NNLO) in the strong coupling. We have checked that the renormalization of rapidity divergences works exactly in the same way as for unpolarized distributions, as it is predicted by the transverse momentum dependent factorization theorem. The present calculation has  a  structure similar to the one  of the  NNLO matching of unpolarized TMDs made in \cite{Echevarria:2016scs}. In the article, we present only the finite part of the coefficient functions, while the logarithmic part can be restored with the renormalization group equations. The full result, including the logarithmic part, is also reported in the attached {\it Mathematica} file.

In the case of transversity, we have considered both the TMDPDF and the TMDFF cases. We have found several analogies and identities between the matching coefficients of transversity and unpolarized distributions, that can serve as a cross-check of both results. The matching coefficients for transversity (given in eqs.~(\ref{C0},~\ref{C1},~\ref{C2}) and eq.~(\ref{C0frag}, \ref{C0z}, \ref{C0zz})) can be readily used in phenomenological applications. A recent review of the phenomenology of transversity in fragmentation can be found in \cite{Metz:2016swz}. Our result is the first calculation of the NNLO matching for transversely polarized TMD operator. To our knowledge the NLO matching coefficient for transversity TMDFF (\ref{c1frag})) is also calculated here for the first time. We stress that this is also the first NNLO evaluation of the matching for a polarized TMD distribution. Therefore, given the result of this work, the transversity TMD distribution is known to the same perturbative order as unpolarized distributions. This fact is important to establish phenomenologically the universality of TMD evolution.

For the pretzelosity, we have found that the expected two-loop matching coefficient is actually zero, despite the fact that the matrix element over free quark for pretzelosity distribution is non-zero. It is an unexpected result, since the analogous quadrapole distribution in the gluon sector (namely linearly polarized TMD gluon distribution) has a non-zero matching already at one-loop level. We have also checked that the LO of large-$N_f$ expansion (given by diagrams with an arbitrary number of fermion bubble insertions, for details see \cite{Scimemi:2017etj}) is also null.  Although these facts do not demonstrate completely that the twist-2 part of pretzelosity is zero at all orders in perturbation theory, certainly they are an evidence of this statement. We conjecture that the pretzelosity distribution does not match the twist-2 distribution, and thus has the leading matching only at the twist-4 level. At the moment we have not found an argument to justify this fact beyond the present calculation. Nevertheless, it agrees with the phenomenological and experimental results that suggest a highly suppressed pretzelosity distribution \cite{Lefky:2014eia,Parsamyan:2018evv}. We have performed the calculation of pretzelosity only for the case of TMDPDF, nonetheless, we expect the same result for the TMDFF.

\subsection*{Acknowledgments}  
D.G.R. and I.S. are supported by the Spanish MECD grant FPA2016-75654-C2-2-P and the group UPARCOS. D.G.R. acknowledges the support of the Universidad Complutense de Madrid through the predoctoral grant CT17/17-CT18/17.
\appendix

\bibliography{TMD_ref}
\bibliographystyle{JHEP}

\end{document}